\documentclass[aps, prd, twocolumn, superscriptaddress,tightenlines, preprintnumbers,nofootnoteinbib]{revtex4-1}
\usepackage[T1]{fontenc}
\usepackage{microtype}
\usepackage[textsize=scriptsize,backgroundcolor=red!70,linecolor=red]{todonotes}
\usepackage{booktabs}
\usepackage{dcolumn}
\usepackage{amsmath}
\usepackage{amsfonts}
\usepackage{amssymb}
\usepackage{graphicx}
\usepackage{hyperref}
\usepackage[all]{hypcap}
\usepackage{paralist}
\usepackage{multirow}

\newcommand{\chisq}{\ensuremath{\chi^2}}

\usepackage{graphicx}% Include figure files
\usepackage{dcolumn}% Align table columns on decimal point
\usepackage{bm}% bold math
\usepackage{epsf}% Include figure files
\usepackage{dcolumn}% Align table columns on decimal point
\def\be{\begin{equation}}
\def\ee{\end{equation}}
\def\bea{\begin{eqnarray}}
\def\eea{\end{eqnarray}}
\def\gsim{\ \rlap{\raise 2pt\hbox{$>$}}{\lower 2pt \hbox{$\sim$}}\ }
\def\lsim{\ \rlap{\raise 2pt\hbox{$<$}}{\lower 2pt \hbox{$\sim$}}\ }
\def\dslash{\kern-4pt \not{\hbox{\kern-2pt $\partial$}}}
\def\pslash{\not{\hbox{\kern-2pt p}}}

\def\beq{\begin{equation}}
\def\eeq{\end{equation}}

\begin{document}
% \ifpdf
%\DeclareGraphicsExtensions{.pdf,.jpg,.mps,.png}
% \else
\DeclareGraphicsExtensions{.eps,.ps}
% \fi

%\preprint{ROME1-1364-2003}

\title{Coherent neutrino scattering and the Migdal effect on the quenching factor}


%%%%%%%%%%%%%%%%%%%%%%%%%%%%%%%%%%%%%%%%%%%%%%%%%%%%%

\author{Jiajun Liao}
\email[Email Address: ]{liaojiajun@mail.sysu.edu.cn}
\affiliation{School of Physics, Sun Yat-sen University, Guangzhou, 510275, China}

\author{Hongkai Liu}
\email[Email Address: ]{hol42@pitt.edu}
\affiliation{Department of Physics and Astronomy, University of Pittsburgh, Pittsburgh, PA 15260, USA}
 
\author{Danny Marfatia}
\email[Email Address: ]{dmarf8@hawaii.edu}
\affiliation{Department of Physics and Astronomy, University of Hawaii at Manoa, Honolulu, HI 96822, USA}

%\collaboration can be followed by \email, \homepage, 
%\thanks as well.
%\date{\today}
%%%%%%%%%%%%%%%%%%%% abstract %%%%%%%%%%%%%%%%%%%%%%%%%%%%%%%%%%%%%%%%%%
\begin{abstract}

Recent measurements of the germanium quenching factor deviate significantly from the predictions of the standard Lindhard model for nuclear recoil energies below a keV. This departure may be explained by the Migdal effect in neutron scattering on germanium.
 We show that the Migdal effect on the quenching factor can mimic the signal of a light $Z^\prime$ or light scalar mediator in coherent elastic neutrino-nucleus scattering experiments with reactor antineutrinos. It is imperative that the quenching factor of nuclei with low recoil energy thresholds be precisely measured close to threshold to avoid such confusion. This will also help in experimental searches of light dark matter.
\end{abstract}
%{11.30.Er.,11.30.Cp.,14.60.Pq,13.15.+g}
%{11.30.Er,11.30.Pb,12.60.Jv}
%\keywords{}
%%%%%%%%%%%%%%%%%%%%%%%%%%%%%%%%%%%%%%%%%%%%%%%%%%%%%%%%%%%%%%%%%%%%%%%%%%
%\maketitle must follow title, authors, abstract, 
%\pacs, and \keywords
\maketitle

%-----------------------------------------------------------%
%%%%%%%%%%%%%%%%%%% SECTION 1 : introduction   %%%%%%%%%%%%%%
%-----------------------------------------------------------%
%\section{Introduction}
%-----------------------------------------------------------%
{\bf Introduction.} 
Coherent elastic neutrino-nucleus scattering (CE$\nu$NS) is a standard model (SM) process in which low-energy neutrinos scatter off the atomic nucleus as a whole via the neutral-current of SM weak interactions~\cite{Freedman:1973yd}. CE$\nu$NS was first observed by the COHERENT experiment in 2017 with a cesium-iodide (CsI) detector using neutrinos produced by stopped pion decay at the Spallation Neutron Source at the Oak Ridge National Laboratory~\cite{Akimov:2017ade}. The observation of CE$\nu$NS opens a new window to probe new physics beyond the SM at low energies~\cite{review}. Recently, CE$\nu$NS has also been measured in an argon detector by the COHERENT collaboration at more than 3$\sigma$ significance~\cite{Akimov:2020pdx}. In addition, the CONNIE~\cite{Aguilar-Arevalo:2019jlr} and CONUS~\cite{Bonet:2020awv} experiments have constrained CE$\nu$NS with reactor antineutrinos in a silicon and germanium detector, respectively. In the near future, more data from the CE$\nu$NS experiments will enable a precision test of the SM at the low energy frontier. 

\begin{figure}
	%\captionsetup{singlelinecheck=on}
	%\centering
	\includegraphics[width=0.45\textwidth]{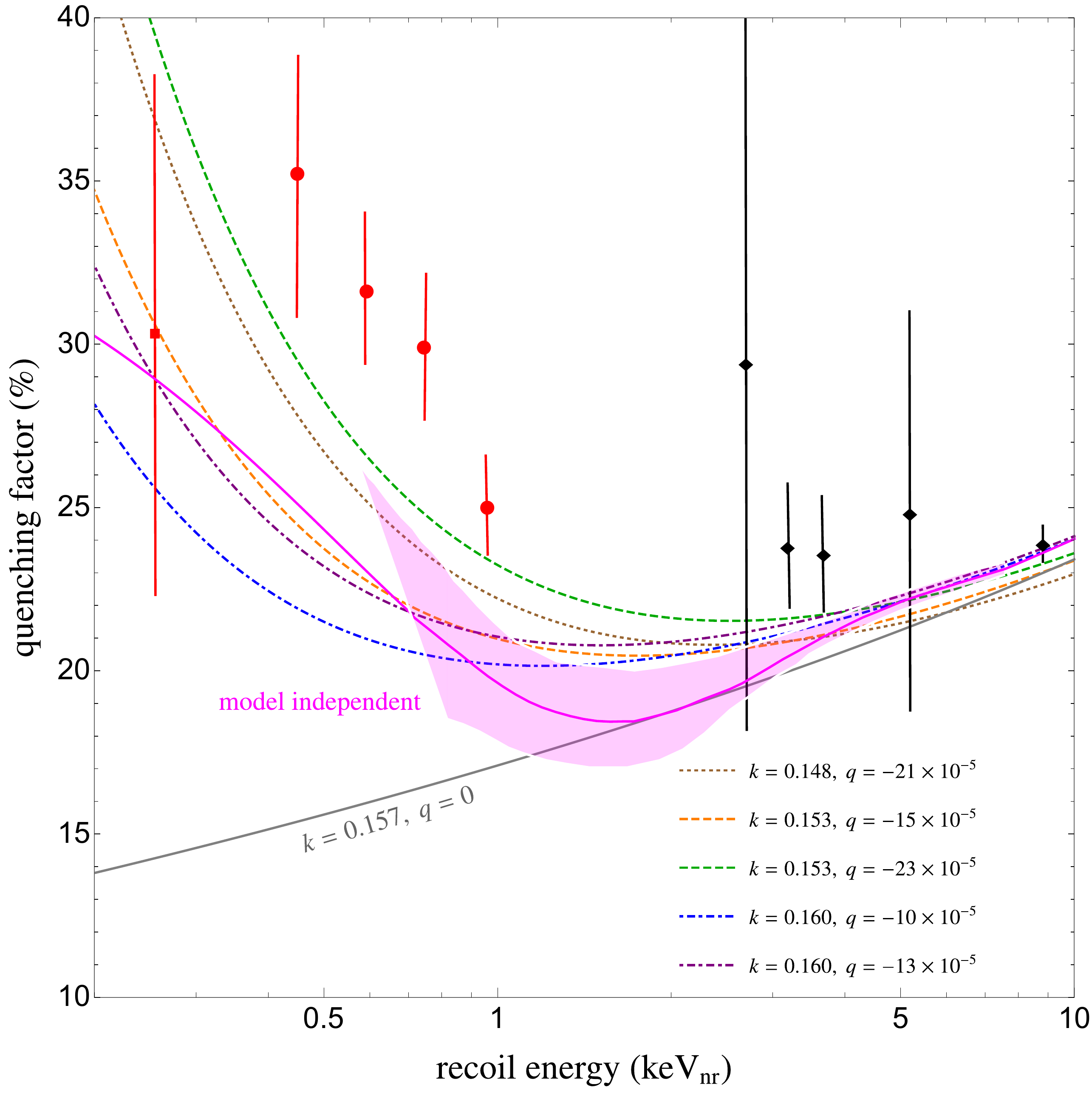}
	\caption{Germanium quenching factor as a function of recoil energy. The new measurements from Ref.~\cite{Collar:2021fcl} are shown in red. The black diamonds are other relatively recent measurements~\cite{texono2,Messous}. The gray solid line corresponds to the standard Lindhard model with $k=0.157$ for germanium. The shaded band is the 95\% C.L. region for the model-independent fit  to the $^{88}$Y/Be$-^{88}$Y/Al residual counts in Ref.~\cite{Collar:2021fcl}. The pink curve outside the band is our extrapolation. The other curves correspond to the modified Lindhard model with various values of the Migdal parameters $(k, q)$.}
	\label{fig:QF}
\end{figure}

Measurements of CE$\nu$NS strongly depend on the quenching factor $Q$, which is defined as the ratio of the observable nuclear recoil energy $E_R$ to that of an electron recoil of the same kinetic energy.  If energy $E_I$ is deposited in the form of ionization or scintillation, $Q\equiv E_I/E_R$. 
%The quenching factor is usually defined as the ratio of the amount of electronic excitation energy produced by nuclear recoils with respect to a recoiling electron of the same energy. 
For $E_R\gtrsim 5\text{ keV}_{nr}$, experimental measurements of the quenching factor agree well with the predictions of the Lindhard model~\cite{Lindard}. However, for sub-keV nuclear recoils, the quenching factors are not well modeled due to uncertainty in the nuclear scattering and stopping at very low energies~\cite{Lindard2, Sorensen:2014sla}.
% and so their values have to rely on experimental measurements. 
Recently, new measurements of the germanium quenching factor have been obtained by using multiple techniques~\cite{Collar:2021fcl}. As Fig.~\ref{fig:QF} shows, the new set of data deviate significantly from the standard Lindhard model for nuclear recoil energies below $\sim 1 \text{ keV}_{nr}$~\cite{Collar:2021fcl}. A model-independent fit  to the $^{88}$Y/Be$-^{88}$Y/Al residual counts in Ref.~\cite{Collar:2021fcl} can only partially explain the low energy excess. More interestingly, the new dataset can be explained by the Lindhard model supplemented with the Migdal effect~\cite{Collar:2021fcl}.

The Migdal effect is the atomic ionization and excitation caused by the displacement between the atomic electrons and the instantaneously recoiling atomic nucleus~\cite{Migdal}. A calculation of the atomic Migdal effect in dark matter direct detection and CE$\nu$NS experiments is presented in Ref.~\cite{Ibe:2017yqa}. A study of the Migdal effect in semiconductors, which accounts for multiphonon production, can be found in Ref.~\cite{Knapen:2020aky}.
The Migdal effect may also affect measurements of the quenching factor which rely on nuclear recoils due to neutron scattering on a nucleus. This in turn, will modify measurements of the CE$\nu$NS spectrum. If CE$\nu$NS experiments use the Lindhard model of the quenching factor, an event excess caused by the Migdal effect may be misidentified as a signal of new physics. 
In this article, we illustrate how the Migdal effect on the quenching factor can mimic new physics signals in a CE$\nu$NS experiment that uses reactor antineutrinos and a germanium detector.

{\bf Migdal effect and the modified Lindhard model.}
The prevailing theoretical prediction of the quenching factor is given by the Lindhard  model~\cite{Lindard}:
\begin{align}
Q(E_R)&= \frac{k\,g(\epsilon)}{1+k\,g(\epsilon)}\,,
\label{eq:lindhard1}
\\
{\text{where}}\  g(\epsilon) &= 3\,\epsilon^{0.15} + 0.7\,\epsilon^{0.6}+\epsilon\,
\\
{\text{with}} \  \epsilon &= 11.5\,Z^{-\frac{7}{3}}\,E_R\,.
\end{align}
Here, $Z$ is the atomic number of the recoiling nucleus, $\epsilon$ is a dimensionless parameter, $E_R$ is the nuclear recoil energy in keV, and $k$ quantifies the electronic energy loss. In the standard Lindhard theory, 
%$k =0.133 Z^{\frac{2}{3}} A^{-\frac{1}{2}} =0.157 $ for germanium.
$k \approx0.157 $ for germanium.

The standard Lindhard model was modified in Ref.~\cite{Sorensen:2014sla} by introducing an additional term to Eq.~(\ref{eq:lindhard1}):
\begin{align}
\label{mod}
Q(E_R)&= \frac{k\,g(\epsilon)}{1+k\,g(\epsilon)}-{q\over \epsilon}\,,
\end{align}
where $q$ is a negative (positive) parameter if the energy given to electrons is enhanced (cutoff).  Note that the introduction of $q$ modifies the quenching factor at low energies and approaches the standard Lindhard model at high energies. From Fig.~\ref{fig:QF}, we see that the germanium quenching factor data favor a negative value of $q$. Although uncertainties in the quenching factor are large~\cite{Lin:2007ka}, the standard Lindhard model (with $q=0$) cannot produce the sharp upturn in the quenching factor at low recoil energies.
In Ref.~\cite{Collar:2021fcl}, the Migdal effect has been invoked to explain the upturn, but the integrated ionization probability needed is seven times larger than estimated in Ref.~\cite{Ibe:2017yqa}. Such an enhancement may occur in semiconductor targets relative to atomic targets~\cite{Knapen:2020aky}. 
In what follows, we describe the Migdal effect on the quenching factor by a negative $q$ in the modified Lindhard model. For $q<0$, we will refer to a pair of values of ($k,q)$, as Migdal parameters. We choose the Migdal parameters so that the quenching factor is compatible with measurements at high recoil energies~\cite{Lin:2007ka}, but do not necessarily pass  through the data below 1~$\rm keV_{ee}$. While we are motivated by the observed upturn, we await confirmation of these data by other experiments to take the data as certain.

{\bf CE$\nu$NS.}
In the SM, CE$\nu$NS is induced via the exchange of a $Z$ boson between neutrinos and quarks. The differential cross section of CE$\nu$NS in the SM is given by~\cite{Freedman:1973yd}
\begin{equation}
\label{eq:crossSM}
\frac{d\sigma_{SM}}{dE_R}=\frac{G_F^2M}{4\pi}q_W^2\left(1-\frac{ME_R}{2E_{\nu}^2}\right)F^2(\mathfrak{q})\,,
\end{equation}
where $G_F$ is the Fermi coupling constant, $E_{\nu}$ is the antineutrino energy,  and $q_W = N - (1 -4 \sin^2\theta_W)Z$ is the weak nuclear charge with $\theta_W$ the weak mixing angle. Here, $F(\mathfrak{q})$ is the Klein-Nystrand parameterization of the nuclear form factor given by~\cite{Klein:1999gv}
\beq
F(\mathfrak{q})= \frac{4\pi\rho_0}{A\mathfrak{q}^3}(\sin \mathfrak{q}R - \mathfrak{q}R\cos \mathfrak{q}R)\frac{1}{1+a^2\mathfrak{q}^2},
\eeq
where $A$ is the atomic number for the nucleus, $\mathfrak{q}$ is the momentum transfer, the range of the Yukawa potential $a = 0.7$~fm, the nuclear radius $R = A^{1/3}r_0$ with the proton radius $r_0 = 1.3$~fm, and the nuclear density $\rho_0 = \frac{3}{4\pi r_0^3}$. Note that the form factor parameterization and form factor uncertainties are not important for CE$\nu$NS induced by reactor antineutrinos due to the low momentum transfer involved~\cite{AristizabalSierra:2019zmy}.

The CE$\nu$NS cross section will be modified by a new mediator that couples to neutrinos and quarks. The differential cross section that includes new universal flavor-conserving interactions mediated by a light vector $Z^\prime$ with mass $M_{Z^\prime}$ and coupling $g^\prime$ is~\cite{review}
\begin{equation} 
\label{eq:crossV}
\frac{d\sigma_{SM+Z^{\prime}}}{dE_R}=\left(1-\frac{q_{Z^{\prime}}}{q_W}\right)^2\frac{d\sigma_{SM}}{dE_R}\,,
\end{equation}
with the effective charge $q_{Z^{\prime}}$ given by
\begin{equation}
\label{eq:QLV}
q_{Z^{\prime}} = \frac{3\sqrt{2}\left(N+Z\right){g^{\prime}}^2}{G_F\left(2M E_R + M_{Z^{\prime}}^2\right)}\,.
\end{equation}
The differential cross section that includes new universal flavor-conserving interactions mediated by a light scalar $\phi$ with mass
$M_\phi$ and coupling $g_\phi$
is~\cite{review}
\begin{equation}
\label{eq:crossS}
\frac{d\sigma_{SM+\phi}}{dE_R}=~\frac{d\sigma_{SM}}{dE_R}+ \frac{d\sigma_{\phi}}{dE_R}\,,
\end{equation}
where
\begin{equation}
\frac{d\sigma_{\phi}}{dE_R}~=~\frac{G_F^2}{4\pi}q_{\phi}^2\frac{2ME_R}{E_{\nu}^2}MF^2(\mathfrak{q})\,,
\end{equation}
with $q_{\phi}$ given by
\begin{equation}
\label{eq:QLS}
q_{\phi}~=~\frac{\left(14 N + 15.1 Z\right)g_{\phi}^2}{\sqrt{2}G_F\big(2ME_R + M_{\phi}^2\big)}\,.
\end{equation}

\begin{figure*}[t]
	%\captionsetup{singlelinecheck=on}
	%\centering
	\includegraphics[width=0.45\textwidth]{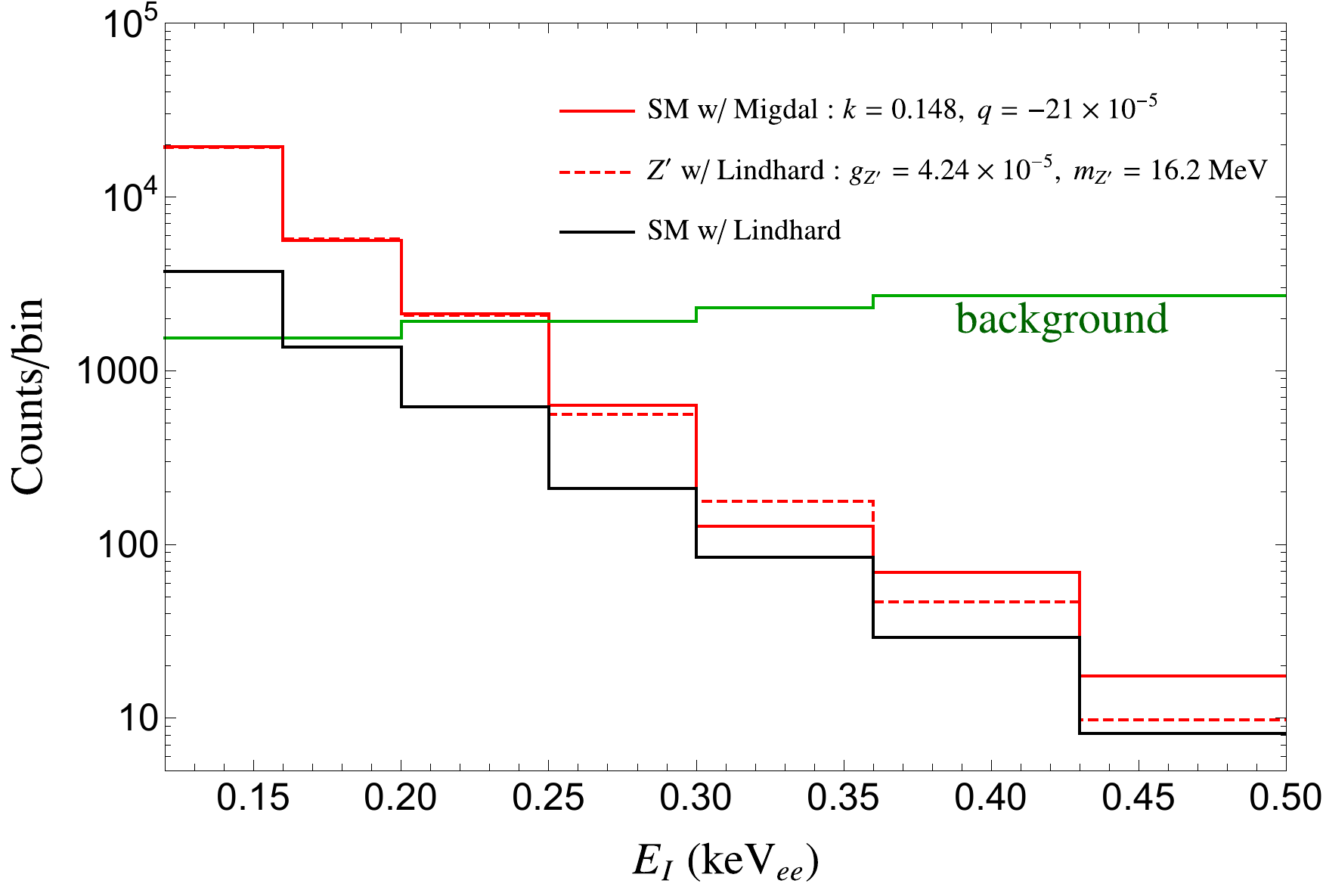}\,\,\,
	\includegraphics[width=0.45\textwidth]{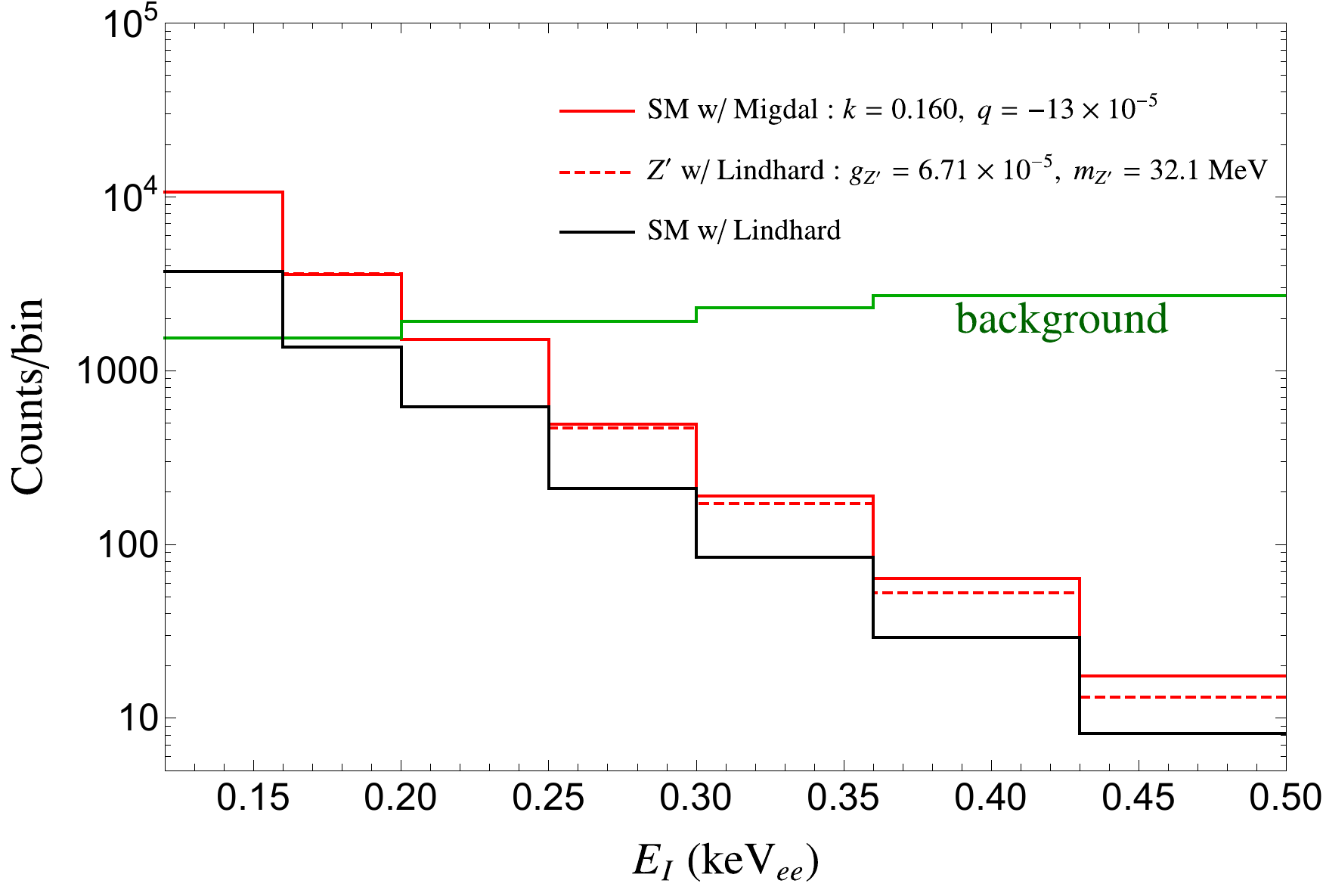}\\
	\includegraphics[width=0.45\textwidth]{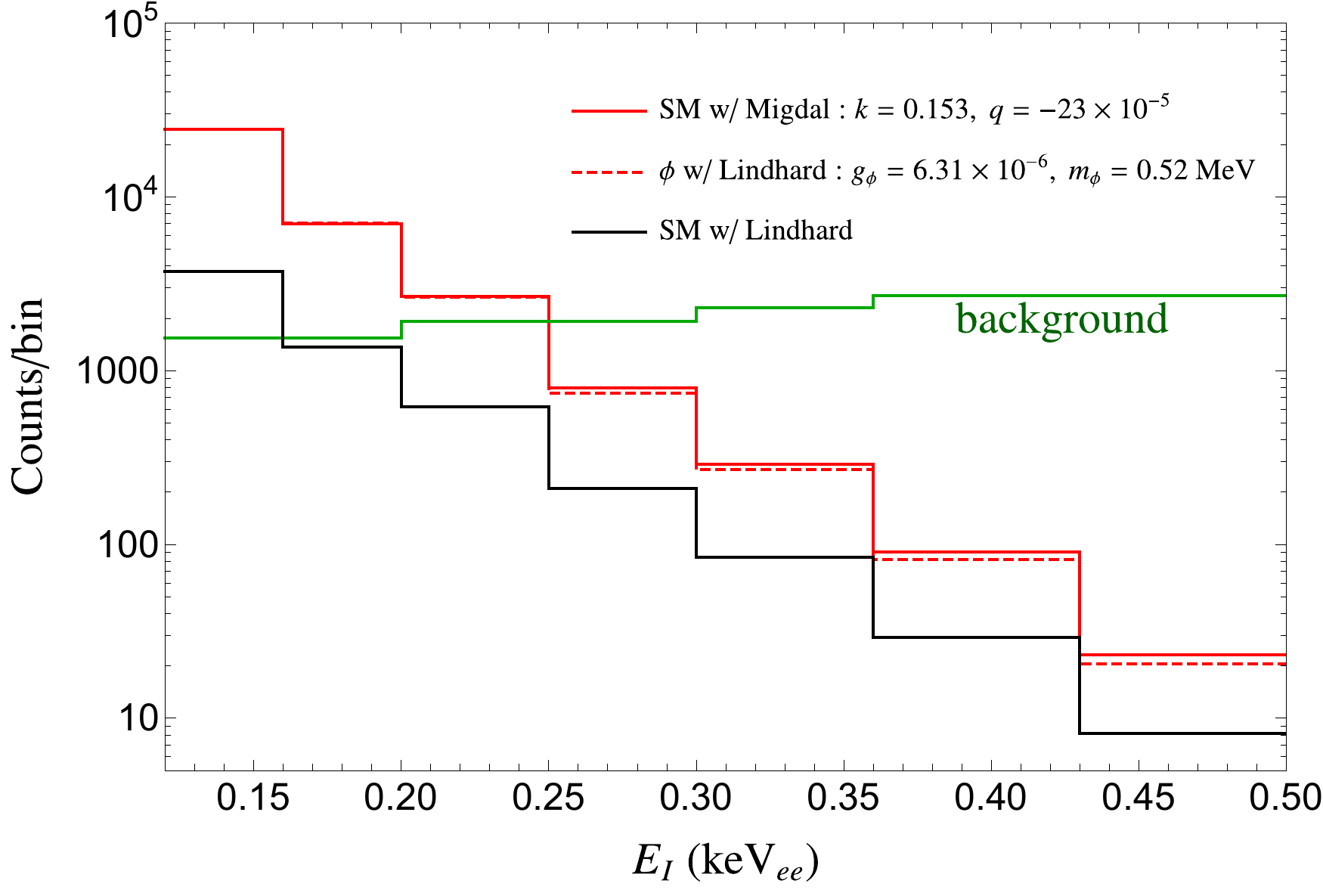}\,\,\,
	\includegraphics[width=0.45\textwidth]{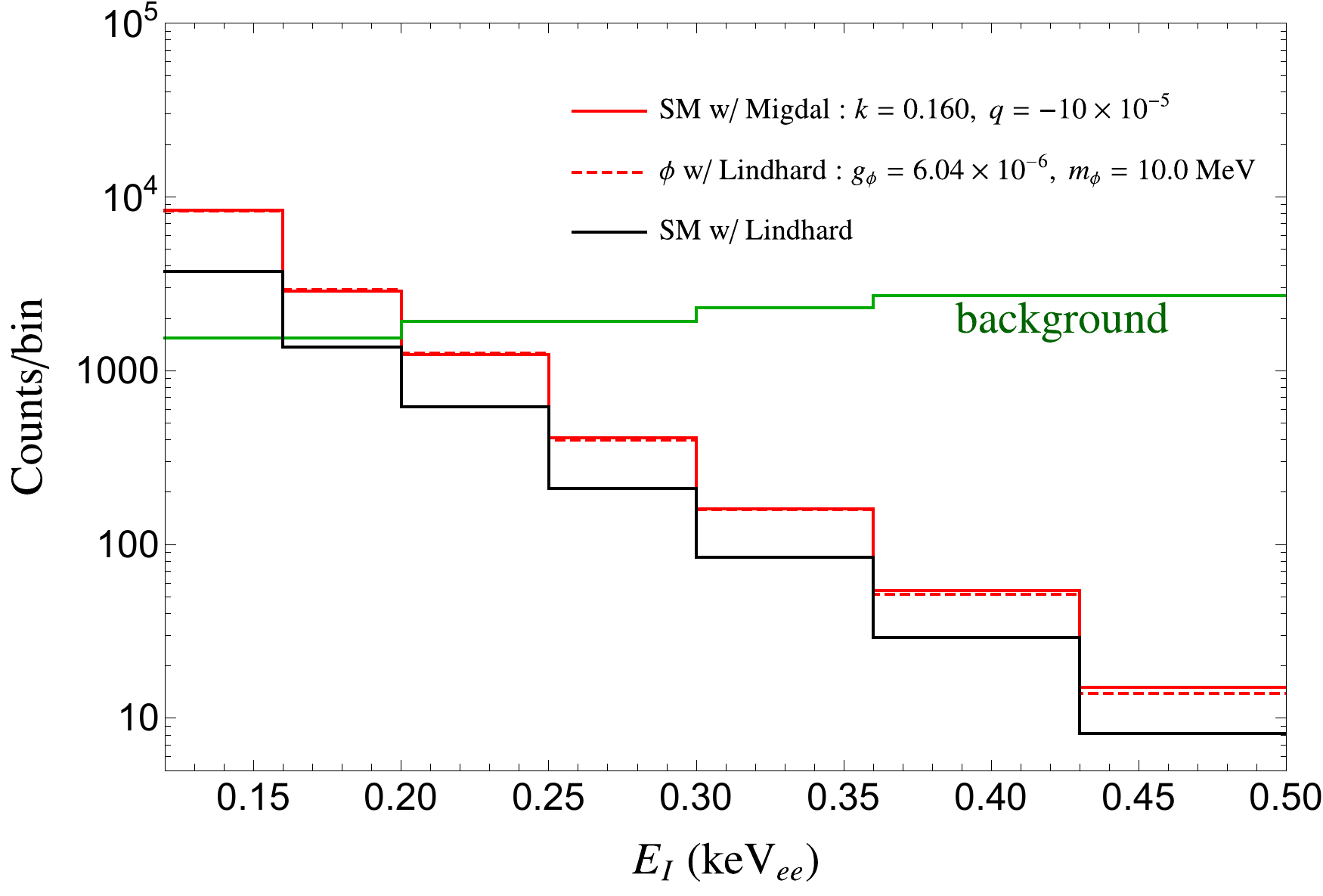}
	\caption{The expected CE$\nu$NS spectra at a germanium detector. 
	The red solid histograms correspond to the SM with the Migdal effect. These spectacularly match the red dashed histograms for the new physics scenarios ($Z^\prime$ and scalar $\phi$ mediators) with the standard Lindhard model. For comparison, SM predictions that assume the standard Lindhard model are shown in black. The green histogram is the background. Note that the lower recoil energy bins are smaller.}
	\label{fig:spectrum}
\end{figure*}

{\bf Experimental setup.}
Since the Migdal effect on the quenching factor becomes significant at low energies, we consider a CE$\nu$NS experiment that utilizes reactor antineutrinos and a p-type point contact high-purity Ge detector that has a very low recoil energy threshold~\cite{Baxter:2019mcx}. The differential CE$\nu$NS event rate is
\beq
\frac{dR}{d E_R} = N_T \int \frac{d\Phi}{dE_\nu}\frac{d\sigma}{dE_R} d E_\nu\,,
\label{eq:eventrate}
\eeq
where $N_T$ is the number of nuclei in the detector. The differential cross section $\frac{d\sigma}{dE_R} $ is given by Eqs.~(\ref{eq:crossV}) and~(\ref{eq:crossS}) for a light $Z^\prime$ and scalar, respectively. 
The reactor antineutrino flux $\frac{d\Phi}{dE_\nu}$ is given by
\beq
\frac{d\Phi}{dE_\nu} = \frac{P}{4\pi d^2 \tilde{\epsilon} } \left(\frac{d N_\nu}{dE_\nu}\right)\,,
\eeq
where $P=3.9$~GW is the reactor thermal power, $d=20$~m is the distance between reactor and detector, and $\tilde{\epsilon}=205.24$ MeV is the average energy released per fission.  The antineutrino spectrum per fission $\frac{d N_\nu}{d E_\nu}$ is taken from Appendix A of Ref.~\cite{Aguilar-Arevalo:2019zme}.

In our analysis, we assume a perfect energy resolution for the detector and identify the measured energy as the ionization energy $E_I$. (The confusion between the Migdal effect and new physics that we seek to demonstrate will only worsen if energy resolution effects are included.) The number of events with measured energy in the $i^{\rm th}$ bin $[E_I^i, E_I^{i+1}]$  is given by~\cite{ Aguilar-Arevalo:2019zme} 
\beq
N_i=t\int_{E_I^i}^{E_I^{i+1}}\eta \frac{dR}{d E_R} \left(\frac{1}{Q}-\frac{E_I}{Q}\frac{dQ}{dE_I} \right)dE_I\,,
\label{eq:counts}
\eeq
where $t=7$~kg$\cdot$year is the exposure time, and $\eta=80$\% is the signal efficiency~\cite{Baxter:2019mcx}. We assume the high purity germanium isotope in the detector is $^{72}$Ge.  Our spectra have 7 bins from 0.12 to 0.5 $\rm{keV}_{ee}$ with the width of each bin taken to be twice the energy resolution given in Ref.~\cite{Baxter:2019mcx}. The background is taken to be 15 counts/$\rm keV_{ee}$/kg/day (ckkd).
%The spectra for both the SM and new physics predictions are shown in Fig.~\ref{fig:spectrum}. 

In our simulations of the CE$\nu$NS spectrum, we neglect ionization events produced by the Migdal effect that appear as unquenched electronic recoils. For $E_I \sim 0.1\ \rm keV_{ee}$, from Fig.~5 of Ref.~\cite{Bell:2019egg}, we see that  the Migdal event rate from $pep$  and $^7$Be solar neutrinos is $\sim 1$~count/$\rm{keV_{ee}}$/ton/year. Since the flux at the detector of reactor neutrinos of the same energies is 
$\sim 3 \times 10^4$ ($\sim 1500$) times larger than  the $pep$ ($^7$Be) solar neutrino flux at Earth, the corresponding event rate is 
smaller than $0.1$~ckkd. We estimate this rate to be enhanced by a factor of a few for the integrated ionization probabilities that reproduce the quenching factors with $q<0$ in Fig.~\ref{fig:QF}~\cite{Collar:2021fcl}. We treat these electronic recoil events as separable from the nuclear recoil events or as part of the 15~ckkd background.

{\bf Analysis and results.}
%To illustrate the Migdal effect on the quenching factor can be confused with new physics signals in a CE$\nu$NS experiment,
 We now compare the spectra of two scenarios: (i) the light $Z^\prime$ or scalar mediator with the standard Lindhard model for the quenching factor; (ii) the SM with the Migdal effect on the quenching factor parameterized by Migdal parameters.
%We firstly simulate the data in the light vector model with $m_{Z^\prime}=10$ MeV and $g_{Z^\prime}=2.0\times 10^{-5}$ assuming the model-independent fit for the quenching factor. Then we scan the parameter space in the SM with modified Lindhard quenching factor by varying both $k$ and $q$. 
To evaluate the statistical significance of the two scenarios, we define 
\begin{align}
\chisq = \sum_{i=1}^{7} 2(\alpha N_{i}^{th} - N_{i}^{exp} + N_{i}^{exp}\ln\frac{N_{i}^{exp}}{\alpha N_{i}^{th}})+\frac{(1-\alpha)^2}{\sigma_\alpha^2} \,,\nonumber
\end{align}
where $\sigma_\alpha=5\%$ is the percent uncertainty in the reactor neutrino flux normalization, $N_{i}^{exp}$ is the simulated event counts per bin, and $N_{i}^{th}$ is the expected number of events per bin calculated using Eq.~(\ref{eq:counts}). 
We find that the Migdal effect on the quenching factor can be easily mimicked by a light $Z^\prime$ or a scalar mediator; see Fig.~\ref{fig:spectrum}. 
The minimum $\chi^2$ (with 4 degrees of freedom) for these scenarios are listed in Table~\ref{Table: chi2}. 

\begingroup
\setlength{\tabcolsep}{10pt} % Default value: 6pt
\renewcommand{\arraystretch}{1.} % Default value: 1
\begin{table}
	\centering
	\begin{tabular}{c |  c|  c|  c|c}
		\toprule
		$m_{Z^\prime}$/MeV  & $g_{Z^\prime}\times10^{5}$ & $k$  & $q\times10^5$  &$\chi^2_\text{min}$\\
		\midrule
		$16.2$ & $4.24$ & $0.148$ & $-21$ &  5.95\\
		\midrule
		$24.6$& $5.40$ & $0.153$ & $-15$ &1.32\\ 
		\midrule
		$32.1$& $6.71$ & $0.160$ & $-13$ & 0.81\\
		\toprule
		\toprule
		$m_{\phi}$/MeV&$g_{\phi}\times10^{6}$  & $k$  & $q\times10^5$ & $\chi^2_\text{min}$\\
		\midrule
		$0.52$& $6.31$ & $0.153$ & $-23$ &  2.11\\
		\midrule
		$5.0$& $5.52$ & $0.153$ & $-15$ & 0.32\\
		\midrule
		$10.0$& $6.04$ & $0.160$ & $-10$ & 1.46\\
		\toprule
	\end{tabular}
	\caption{Sample $Z^\prime$ and scalar $\phi$ parameters that mimic the SM with the Migdal effect parametrized by the $k$ and $q$ values indicated; the quenching factors are plotted in Fig.~\ref{fig:QF}. The minimum 
		$\chi^2$ values for 4 degrees of freedom show how well the spectra match.}
	\label{Table: chi2}
\end{table}
\endgroup

\begin{figure}[t]
	%\captionsetup{singlelinecheck=on}
	\centering
	\includegraphics[width=0.45\textwidth]{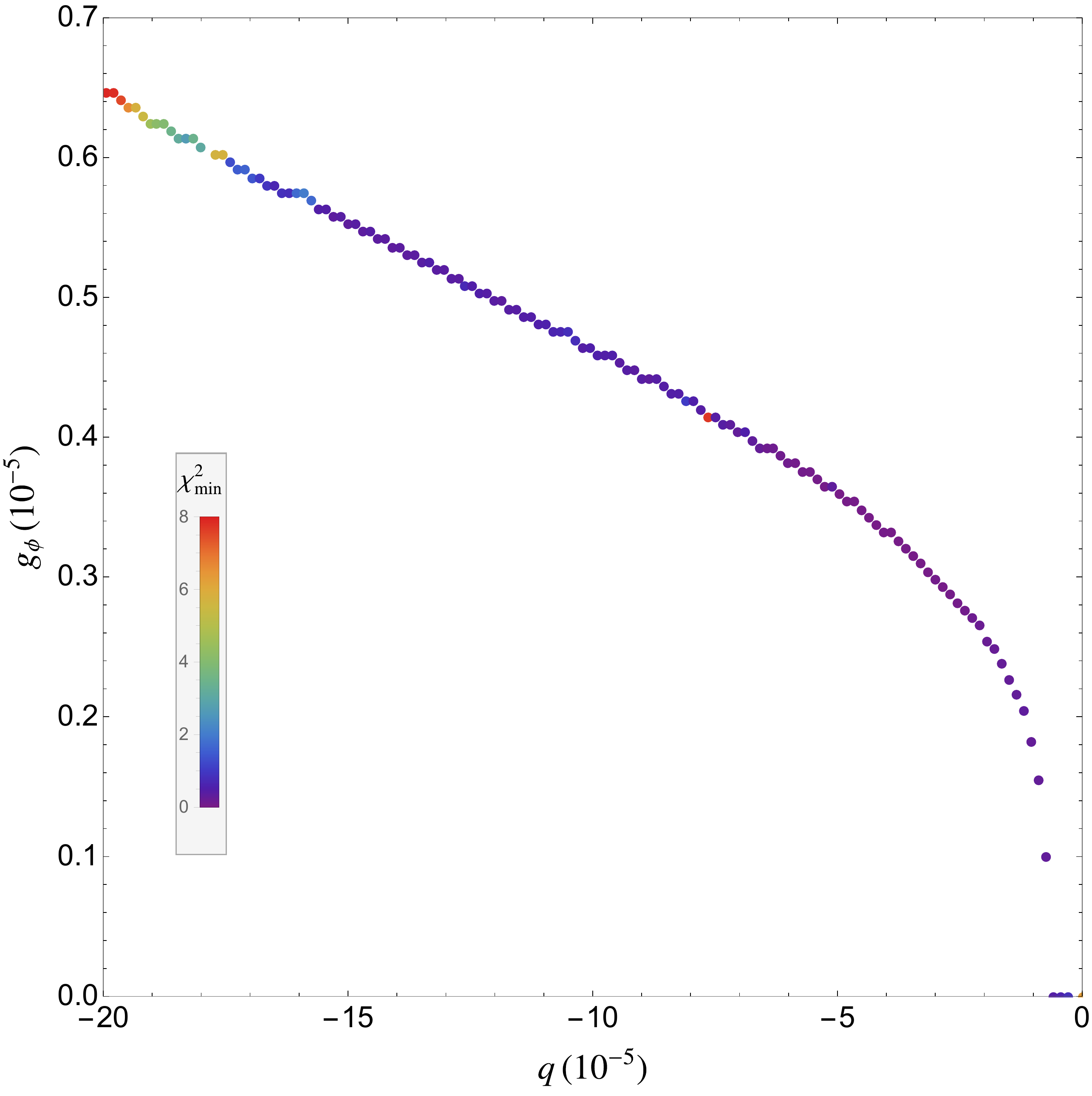}\,\,\,
	\caption{The correlation between $q$ and the coupling constant $g_\phi$ of a scalar mediator of mass $m_\phi=5$~MeV. The colors indicate $\chi^2_{\rm min}$ (for 5 degrees of freedom).
	%Right: The $2\sigma$ allowed regions for the $Z^\prime$  (scalar) model from a fit to simulated SM data with $k=0.153$ and $m_{Z^\prime}=24.6$~MeV ($m_{\phi}=5$~MeV).
	}	\label{fig:corr}
\end{figure}

An illustration of the correlation between $q$ and the coupling constant $g_\phi$ of the scalar mediator of mass $m_\phi=5$~MeV is shown in Fig.~\ref{fig:corr}. We create many SM datasets by fixing $k=0.157$ and varying $q$, and fit the coupling constant (with $k=0.153$) to each dataset; we require $\chi^2_{\rm min}< 8$ for a fit to be considered satisfactory. As expected, the coupling constant that fits the SM data becomes smaller as $|q|$ decreases. As $q \to 0$, also $g_\phi \to 0$, indicating that new physics is not needed. For \mbox{$q \lsim -17\times 10^{-5}$}, the fit worsens, indicating that a value different from $k=0.153$ is preferred in the new physics scenario.

\begin{figure*}[t]
	%\captionsetup{singlelinecheck=on}
	\centering
	\includegraphics[width=0.45\textwidth]{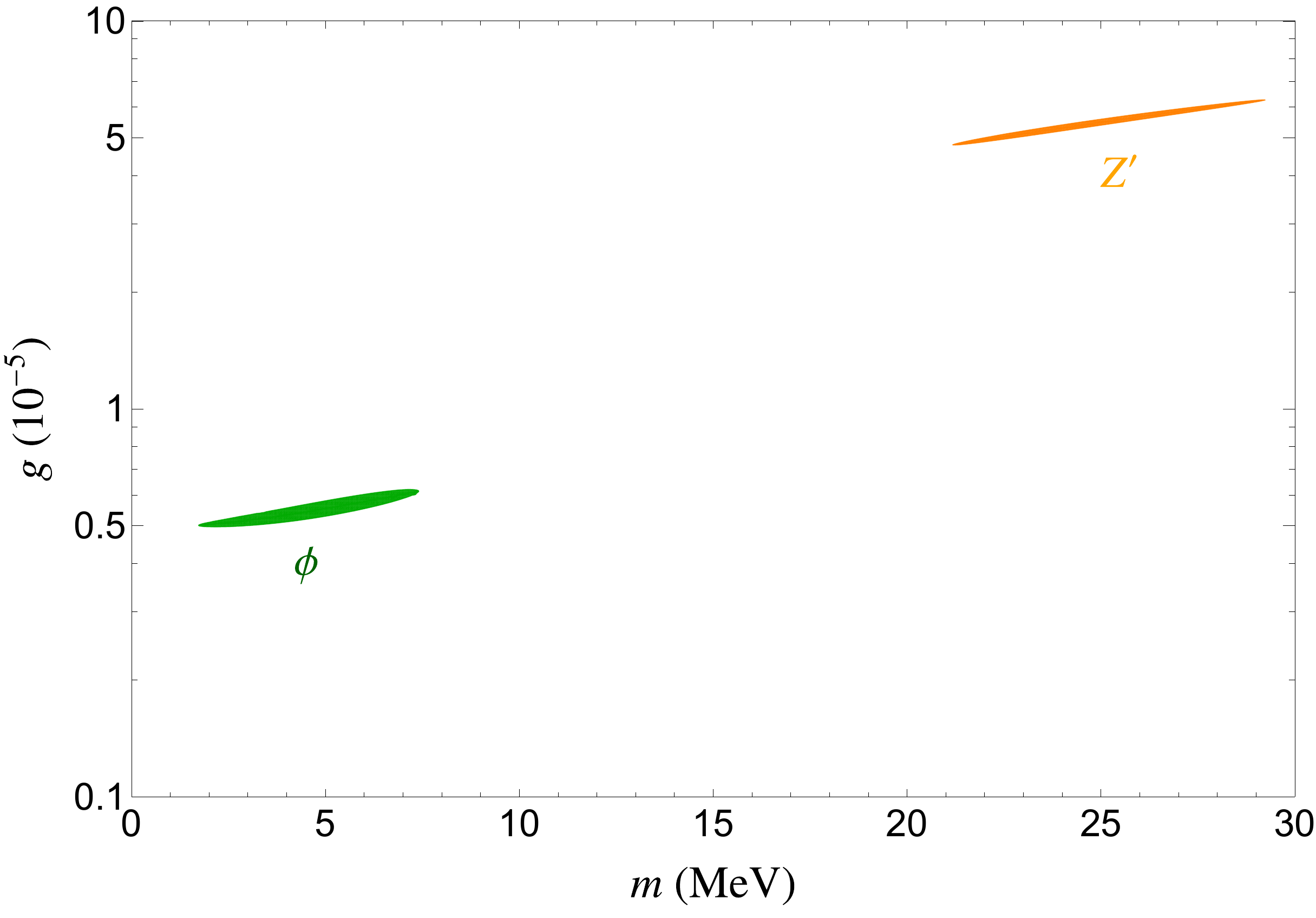}\,\,\,
	\includegraphics[width=0.465\textwidth]{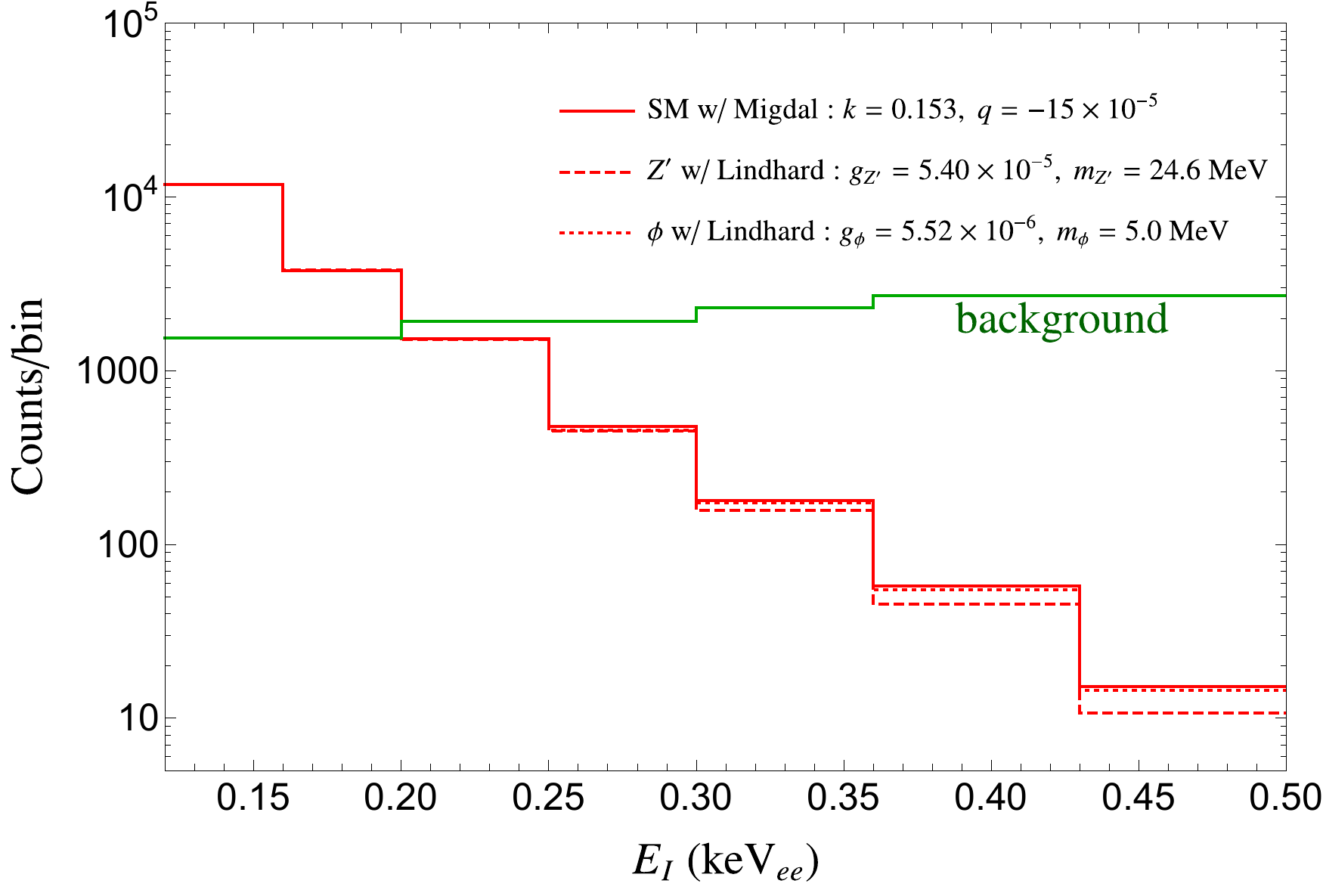}\,
	\caption{Left panel: The $2\sigma$ allowed regions for the light $Z^\prime$ and light scalar models from a fit to simulated SM data with Migdal parameters, $k=0.153$ and $q=-15\times 10^{-5}$. We fit the new physics scenarios with the standard Lindhard model for the quenching factor. Right panel: The spectra for the best-fit points and the SM.
	%Right: The $2\sigma$ allowed regions for the $Z^\prime$  (scalar) model from a fit to simulated SM data with $k=0.153$ and $m_{Z^\prime}=24.6$~MeV ($m_{\phi}=5$~MeV).
	}	\label{fig:countour}
\end{figure*}

Interestingly, we find that both the light $Z^\prime$ and scalar cases (with the standard Lindhard model) can fit the SM spectrum with the Migdal parameters, $k=0.153$ and $q=-15\times 10^{-5}$. This will lead to confusion in determining the nature of new physics.
%if the data is simulated in the SM with Migdal parameter $k=0.153$ and $q=-15\times 10^{-5}$, both the light $Z^\prime$  and scalar cases can fit the data well with the standard Lindhard model. 
To demonstrate this, we simulate SM data assuming the quenching factor is given by the Migdal parameters, $k=0.153$ and $q=-15\times 10^{-5}$, and fit the light $Z^\prime$ and scalar models (with the standard Lindhard model as the quenching factor). The $2\sigma$ allowed regions and the spectra for the best-fit points (in Table~\ref{Table: chi2}) are shown in Fig.~\ref{fig:countour}. %The best-fit point occurred at $m_{Z\prime}=25.1$ MeV and $g_{Z^\prime}=5.5\times 10^{-5}$ ($m_{\phi}=5.0$ MeV and $g_{Z^\prime}=5.5\times 10^{-6}$) with $\chisq_\text{min}=0.43$ for the light $Z^\prime$ case. 
 The two allowed regions show that a light $Z^\prime$ cannot be distinguished from a light scalar mediator  if the standard Lindhard model is assumed for the quenching factor. The spectra in the right panel show that both new physics models cannot be distinguished from the SM with a quenching factor modified by the Migdal effect.

  %We also fit the simulated SM data with varying $q$. The right panel shows that for a fixed mediator mass, the coupling constant that fits the SM data becomes smaller as $|q|$ decreases. This is because the number of events at low recoil energies decreases if $|q|$ decreases.
%for the SM predictions with the same Migdal parameter on the quenching factor.
%
%\begin{figure}[t]
%	%\captionsetup{singlelinecheck=on}
%	%\centering
%	\includegraphics[width=0.45\textwidth]{ZSspecplot.pdf}\\
%	\caption{The expected spectra at the CE$\nu$NS experiment with a germanium detector.}
%	\label{fig:ZS}
%\end{figure}
%

{\bf Summary.} 
The Lindhard model is widely used to describe the quenching factor in CE$\nu$NS and dark matter direct detection experiments. Recent measurements of the quenching factor in germanium indicate a departure from the standard Lindhard model~\cite{Collar:2021fcl}. This deviation can be interpreted as evidence of the Migdal effect on the quenching factor. We parameterized the Migdal effect with a negative value of $q$ in the modified Lindhard model of Eq.~(\ref{mod}). We showed that both light $Z^\prime$ and scalar mediators with the standard Lindhard model can mimic the SM with the Migdal effect on the quenching factor.
In fact, the SM with  a given set of Migdal parameters can be simultaneously degenerate with both
a $Z^\prime$ and a scalar model. To avoid such confusion in detecting new physics at future CE$\nu$NS experiments, a precise measurement of the quenching factor of nuclei with low recoil energy thresholds is urgently required. The precision attainable is as yet unknown. A detailed analysis of how to break the degeneracy is left for future work.

 {\it Acknowledgments.} 
We thank J. Estrada and G. Fernandez for helpful discussions.
J.L. is supported by the National Natural Science Foundation of China under Grant No. 11905299 and Guangdong Basic and Applied Basic Research Foundation under Grant No. 2020A1515011479. H.L. is supported in part by the U.S. Department of Energy under grant No. DE-FG02-95ER40896 and in part by the PITT PACC. D.M. is supported in part by the U.S. Department of Energy under Grant No. de-sc0010504.

%\newpage
%%%%%%%%%%%%%%%%%%%%%%%%%%%%%%%%%%%%%%%%%%%%%%%%%%%%%%%%%%%%

\end{document}